\documentclass[twocolumn, prd, nofootinbib, showpacs, amsmath, amssymb] {revtex4}

\usepackage{graphicx}

\begin{document}

\title{Space-time uncertainty relation and operational definition of dimension}

\author{ Michael~Maziashvili}
\email{maziashvili@iphac.ge} \affiliation{ Andronikashvili
Institute of Physics, 6 Tamarashvili St., Tbilisi 0177, Georgia }

\begin{abstract}

Operational definition of space-time in light of quantum mechanics
and general relativity inevitably indicates an intrinsic
imprecision in space-time structure which has to do with
space-time dimension as well. The operational dimension of
space-time turns out to be a scale dependent quantity slightly
smaller than four at distances $\gg l_P$. Close to the Planck
length the deviation of space-time dimension from four becomes
appreciable. The experimental bounds on the deviation of
space-time dimension from four coming from the electron ${\tt g} -
2$ factor, Lamb shift in hydrogen atom and the perihelion shift in
the planetary motion are still far from the theoretical
predictions.

\end{abstract}

\pacs{04.60.-m, 06.20.Dk, 11.10.Kk}




\maketitle

\section*{Introduction}

From the inception of quantum mechanics the concept of measurement
has proved to be a fundamental notion for revealing a genuine
nature of physical reality \cite{Heisenberg}. The background
space-time representing a frame in which everything takes place is
often taken for granted. One of the most fundamental concepts in
physics is the very notion of space-time dimension. Our
fundamental theories of physics usually do not predict the
space-time dimension. A notable exception is provided by
superstring theory the internal consistency of which demands
space-time dimension to be $10$ \cite{superstring}. Usually we are
accustomed to taking for granted that our space-time is four
dimensional. (Also in considering higher-dimensional possibilities
we invariably have in mind some specific integer value for
dimension). As it was noticed in \cite{ZS} because of finite
resolution of space-time we should expect the operational
definition of dimension to result in non-integer value, somewhat
smaller than four. It was noticed long ago \cite{Mead} that the
basic principles of quantum theory and gravitation immediately
indicate a finite space-time resolution. Namely, let us ask to
what maximal precision can we mark a point in space by placing
there a test particle. Throughout this paper we will assume $\hbar
= c = 1$. In the framework of quantum field theory a quantum takes
up at least a volume, $\delta x^3$, defined by its Compton
wavelength $\delta x \gtrsim 1/m$. Not to collapse into a black
hole, general relativity insists the quantum on taking up a finite
amount of room defined by its gravitational radius $\delta x
\gtrsim l_P^2m$. Combining together both quantum mechanical and
general relativistic requirements one finds
\begin{equation}\label{abslimit} \delta x \gtrsim
\mbox{max}(m^{-1},~l_P^2m)~.\end{equation} From this equation one
sees that a particle occupies at least the volume $ \sim l_P^3 $.
Therefore in the operational sense the point can not be marked to
a better accuracy than $ \sim l_P^3 $ (this point of view was
carefully analyzed through a number of {\tt Gedankenexperiments}
in \cite{Mead}). Since our understanding of time is tightly
related to the periodic motion along some length scale, this
result implies in general an impossibility of space-time distance
measurement to a better accuracy than $\sim l_P$. Physical meaning
of this limitation consists in significant magnification of the
space-time fluctuations during the observation when the measured
length scale approaches the Planck one. Roughly it happens because
refined length measurement requires large momentum according to
Heisenberg uncertainty principle, but when the momentum becomes
too large its gravitational disturbance of the region under
measurement becomes appreciable. So this discussion is complectly
in the spirit of quantum mechanical philosophy.

In this Letter following the reasoning of paper \cite{ZS} and
taking into account an unavoidable imprecision in space-time
structure \cite{Mead, SW, Karol, Pady, Maggiore, Ahluwalia, AC,
NvD, Sasakura, mazia, Amelino-Camelia, mazia1} we estimate
space-time dimension and consider some of the experimental bounds
on it.

\subsection*{Space-time uncertainty relation}

First let us summarize different approaches for operational
definition of Minkowskian space-time that enables one to estimate
the rate of quantum-gravitational fluctuations of the background
metric. What we are interested in is to quantify to what maximal
precision can we measure the space-time distance for Minkowski
space. For space-time measurement an unanimously accepted method
one can find in almost every textbook of general relativity
consists in using clocks and light signals \cite{FockLL}. Let us
consider a light-clock consisting of a spherical mirror inside
which light is bouncing. That is, a light-clock counts the number
of reflections of a pulse of light propagating inside a spherical
mirror. Therefore the precision of such a clock is set by the size
of the clock. The points between which distance is measured are
marked by the clocks, therefore the size of the clock $2r_c$ from
the very outset manifests itself as an error in distance
measurement. Let us call it a mechanical error and denote by
$\delta l_{mech} \simeq r_c$. Another source of error is due to
quantum fluctuations of the clock. Namely denoting the mass of the
clock by $m$ one finds that the clock is characterized with spread
in velocity

\[\delta v = {\delta p \over m} \sim {1 \over m\, r_c}~,\] and
correspondingly during the time $t = l$ taken by the light signal
to pass the distance $l$ the clock may move the distance $t\delta
v$. In what follows we will refer to it as a quantum error and
denote by $\delta l_{quant} \simeq l/mr_c$. The total uncertainty
in measuring the lengths scale $l$ takes the form
\[\delta l \gtrsim r_c + {l\over m\,r_c}~. \] Minimizing this
expression with respect to the size of clock one finds
\begin{equation}\label{minwitsize}r_c \simeq \sqrt{l\over m}~~\Rightarrow~~~~\delta l \gtrsim
\sqrt{l\over m}~.\end{equation} By taking the mass of the clock to
be large enough the uncertainty in length measurement can be
reduced but one should pay attention that simultaneously the size
of the clock diminishes and its gravitational radius increases.
The measurement procedure to be possible we should care the size
of the clock not to become smaller than its gravitational radius
to avoid the gravitational collapse of the clock into a black
hole. So that there is an upper bound on the clock mass
\[r_c^{min} \simeq \sqrt{{l\over m_{max}}} \simeq l_P^2
m_{max}~,~~\Rightarrow ~~ m_{max} \simeq {l^{1/3}\over
l_p^{2/3}}~,\] which through the equation (\ref{minwitsize})
determines the minimal unavoidable error in length measurement as
\begin{equation}\label{Karol}\delta l_{min} \simeq l_P^{2/3}l^{1/3}
~.\end{equation} This uncertainty relation was first obtained by
K\'arolyh\'azy in 1966 and was subsequently analyzed by him and
his collaborators in much details \cite{Karol}.

Let us refine our consideration by noticing that after introducing
the clock the metric takes the form

\[ds^2=\left(1-{2l_P^2m\over r}\right)dt^2- \left(1-{2l_P^2m\over r}\right)^{-1}dr^2-r^2d\Omega^2~.\]
The time measured by this clock is related to the Minkowskian time
as \cite{FockLL} \[t'=\left(1-{2l_P^2m\over r_c}\right)^{1/2}t~.\]
From this expression one sees that the disturbance of the
background metric to be small, the size of the clock should be
much greater than its gravitational radius $r_c \gg 2l_p^2m$.
Under this assumption for gravitational disturbance in time
measurement one finds \[t'=\left(1-{l_P^2m\over r_c}\right)t~.\]
Thus, in addition to the mechanical and quantum errors we have the
gravitational error as well. The terms contributing to the total
error look like \[\delta l_{mech} \simeq r_c\,,~~~~\delta
l_{quant} \simeq {l \over mr_c}\,,~~~~\delta l_{grav} \simeq
\,l\,{l_P^2m\over r_c}~.\] The mass of the clock can not be less
than $\sim 1/r_c$, it is nothing else but the requirement the
minimal size to be set by the Compton wavelength. On the other
hand to avoid the gravitational collapse into black hole we should
require $r_c \gtrsim l_P^2m$. So what we know on general grounds
is that $1 / r_c \lesssim m \lesssim r_c / l_P^2$. The terms
$\delta l_{quant}$ and $\delta l_{grav}$ depend on the mass.
Minimizing $\delta l_{quant} + \delta l_{grav}$ with respect to
the mass one finds an optimal value of the clock mass to be $m
\simeq m_P$. After this minimization the total error takes the
form \[\delta l \gtrsim r_c + l\,{l_P \over r_c}~.\] This
expression is minimized for the size of the clock $r_c \simeq
(l_Pl)^{1/2}$ determining the minimal unavoidable error in length
measurement as
\begin{equation}\label{Camelia}\delta l_{min} \simeq
(l_Pl)^{1/2}~.\end{equation} This uncertainty relation was
discussed in \cite{Amelino-Camelia}. Compared with
Eq.(\ref{Karol}) this expression implies more imprecision in
length measurement. Notice that if we omit either $\delta
l_{quant}$ or $\delta l_{grav}$ we will arrive at the
Eq.(\ref{Karol}), see for details \cite{mazia1}.

\section*{Random walk approach to the space-time measurement}

To understand the principal features of the random walk approach
it suffices to consider the following one dimensional example (for
a comprehensive review of the random walks one can see
\cite{Chandrasekhar}). Assume a particle undergoes a sequence of
displacements along a straight line in the form of a series of
steps of equal length, $l_s$, each of them being taken with equal
probability in both directions. So that the probability of each
step to be taken either in the forward or in the backward
direction is $1/2$ independently of the direction of all the
preceding steps. After taking $N$ such steps from the origin of
axis the particle could be at any of the points
\[-l_sN\,,~-l_s(N+1)\,,~\cdots\,,-l_s\,,~0\,,~l_s\,,\cdots\,,~l_s(N-1)\,,~l_sN ~.\]
The question we are interested in is to estimate what is the
probability $W(m\,,N)$ that after $N$ displacements the particle
will be at the point $l_sm$. The probability of any given sequence
of $N$ steps is $(1/2)^N$. The required probability $W(m\,,N)$ is
therefore $(1/2)^N$ times the number of distinct sequences of
steps leading to the point $l_sm$ after $N$ steps. In order to
arrive at $l_sm$ among the $N$ steps, we need to make $m$ steps in
the positive direction to reach this point and the remaining steps
$N - m$ should be taken in equal numbers forth and back, that is,
in whole some $( N + m)/2$ steps should be taken in the positive
direction and the remaining ones $( N - m)/2$ in the negative
direction. The number of such distinct sequences can be easily
estimated by observing that for a given sequence the permutations
among $( N + m)/2$ forth and $( N - m)/2$ back elements do not
produce new sequences. Thus for the number of distinct sequences
one finds \[ {N!\over \left[{N-m \over 2} \right]!\left[{N+m \over
2} \right]! } ~.\] Hence \begin{equation}\label{Bernoulli}
W(m\,,N) = {N!\over \left[{N-m \over 2} \right]!\left[{N+m \over
2} \right]! } \left({1\over 2}\right)^N ~.\end{equation}

Now imagine we are measuring some length scale by the ruler. The
ratio of the length scale under measurement to the length of the
ruler determines the number of steps, $N$, we need to perform for
this measurement. Our ruler has some precision, $l_s$, determining
the uncertainty in each measurement. It is natural to assume that
in making $N$ measurements this uncertainty adds up randomly, that
is, during each step the uncertainty is expected to take on $\pm$
sign with equal probability. Hence, under this assumption one
finds that the probability to make the error $l_sm$ in length
measurement by the ruler $N$ times smaller than this length, is
given by the Bernoulli distribution (\ref{Bernoulli}). With
respect to this distribution one can estimate $\langle m^2 \rangle
= N$ \cite{Chandrasekhar}, which determines the mean square
uncertainty in the measurement \[\mbox{uncertainty in the
measurement } = l_s\sqrt{N}~.\] Gravitational field is described
in terms of space-time metric, that is, figuratively speaking it
measures space-time distances. To measure the space-time distance
gravitational field has the only intrinsic length scale $l_P$. If
we assume our ruler is just $l_P$, that is, both its length and
precision are given by the Planck length we arrive at the
Eq.(\ref{Camelia}) \[\delta l \simeq l_P \left({l \over l_P}
\right)^{1/2} = (l_Pl)^{1/2}~.\] So it seems natural to hold that
the gravitational field operating with the Planck length precision
knows space-time distances with the accuracy given by
Eq.(\ref{Camelia}).

\section*{New light from an effective quantum field theory}

Interestingly enough both of the equations (\ref{Karol}) and
(\ref{Camelia}) were derived in the framework of an effective
quantum field theory in \cite{CKN}. The Eq.(\ref{Karol}) emerges
as a relation between UV and IR scales in the framework of an
effective quantum field theory satisfying the black hole entropy
bound. For an effective quantum field theory in a box of size $l$
with UV cutoff $\Lambda$ the entropy $S$ scales as, $S \sim
l^3\Lambda^3$. That is, an effective quantum field theory
 counts the degrees of freedom simply as the numbers of cells
 $\Lambda^{-3}$ in the box $l^3$. Nevertheless, considerations involving black holes demonstrate
that the maximum entropy in a box of volume $l^3$ grows only as
the area of the box \cite{Bekenstein} \[S_{BH} \simeq
\left({l\over l_P}\right)^2~.\] So that, with respect to the
Bekenstein bound \cite{Bekenstein} the degrees of freedom in a
volume should be counted by the number of surface cells $l_P^2$. A
consistent physical picture can be constructed by imposing a
relationship between
 UV and IR cutoffs \cite{CKN}
\begin{equation}\label{BHbound}l^3 \Lambda^3 \lesssim S_{BH}
\simeq  \left({l\over l_P}\right)^2~.
\end{equation} Consequently one arrives at the conclusion that the length $l$,
which serves as an IR cutoff, cannot be chosen independently of
the UV cutoff, and scales as $\Lambda^{-3}$. Rewriting  this
relation wholly in length terms, $\delta l \equiv \Lambda^{-1}$,
one arrives at the Eq.(\ref{Karol}). Is it an accidental
coincidence? Indeed not. The relation (\ref{BHbound}) can be
simply understood from the Eq.(\ref{Karol}). The IR scale $l$ can
not be given to a better accuracy than $\delta l \simeq
l_P^{2/3}\,l^{1/3}$. Therefore, one can not measure the volume
$l^3$ to a better precision than $\delta l^3 \simeq l_P^2\,l$ and
correspondingly maximal number of cells inside the volume $l^3$
that may make an operational sense is given by $(l/l_P)^2$. Thus
the K\'arolyh\'azy relation implies the black-hole entropy bound
given by Eq.(\ref{BHbound}).

From the preceding sections one may find more motivated to use the
Eq.(\ref{Camelia}) instead of Eq.(\ref{Karol}). Again the
effective quantum field theory can help us to gain new insights
into the problem \cite{CKN}. An effective field theory that can
saturate Eq.(\ref{BHbound}) may include many states with
gravitational radius much larger than the box size. To see this,
note that a conventional effective quantum field theory is
expected to be capable of describing a system at a temperature $T$
provided that $T\leq \Lambda$. So long as $T\gg 1/l$, such a
system has thermal energy $M\sim l^3 T^4$ and entropy $S\sim l^3
T^3$.  When Eq.(\ref{BHbound}) is saturated, at $T\sim
(m_P^2/l)^{1/3}$, the corresponding gravitational radius for this
system is $ \sim l (l m_P)^{2/3}\gg l$. That is, since the maximum
energy density in the effective theory is $\Lambda^4$, the
gravitational radius associated with the maximum energy of the
system, $M_{max}\sim l^3\Lambda^4 ~\Rightarrow ~r_g\sim
l_P^2l^3\Lambda^4$, will be greater than the size of the system,
$l$, if UV cutoff is defined from the Eq.(\ref{BHbound}). To be on
the safe side, one can impose stronger constraint requiring the
size of the system to be greater than the gravitational radius
associated to the maximum energy of the system \cite{CKN}
\begin{equation}
\label{blueberunium} l_P^2l^3 \Lambda^4 \lesssim l \, .
\end{equation} Here the IR cutoff scales like $\Lambda^{-2}$. This relation
written in
 length
terms ($\delta l \equiv \Lambda ^{-1},~l_P\equiv m_P^{-1}$) is the
 Eq.(\ref{Camelia}).

So we see what is the effective quantum field theory picture
behind the Eqs.(\ref{Karol}\,,\,\ref{Camelia}) and interplay
between them.

\section*{Space-time dimension}

The general mathematical concept of dimension was put forward long
ago by Hausdorff \cite{Hausdorff}. Let us briefly recall the
definition of  Hausdorff dimension. We have a metric space denoted
by $(\Omega,\,\xi)$, where $\Omega$ stands for the set of points
and $\xi$ denotes the distance (metric) on it. Let $O$ be the
family of all open sets in $\Omega$ and $O_{\epsilon}$ a subset of
it such that \[O_{\epsilon} = \{U \in O\,|\,d(U) \leq \epsilon
\}~,\]

where $d(U)$ is the diameter of $U$ \[d(U) = \sup\{\xi(x,\,y)\,|\,
x\,,y \in U \}~.\] Let us define for an arbitrary subset $E
\subset \Omega$ covered by the countable number of $U_n$th the
following quantity

\[\mu^{\alpha}(E,\,\epsilon) = \inf\left\{\sum\limits_nd(U_n)^{\alpha}\,|\,U_n\in
O_{\epsilon},\,E \subset \bigcup\limits_nU_n\right\}~,\] where the
{\tt infimum} is over all countable open covers of $E$. Then the
Hausdorff dimension is defined as \[ \dim(E)\equiv \alpha_H  =
\sup\{\alpha \geq 0\,|\,\mu^{\alpha}(E)  = \infty  \}~,\] where \[
\mu^{\alpha}(E) = \lim\limits_{\epsilon\rightarrow
0}\mu^{\alpha}(E,\,\epsilon)~. \] This definition is based on an
important property that $\mu^{\alpha}(E) = \infty$ for $\alpha <
\alpha_H$ and $\mu^{\alpha}(E) = 0$ for $\alpha > \alpha_H$. It is
important to notice that the definition of dimension implies the
limit $\epsilon \rightarrow 0$. In any real or {\tt Gedanken}
measurement one always deals with a finite resolution that
naturally leads to the necessity of operational definition of the
space-time dimension \cite{ZS}. From the space-time uncertainty
relation we infer that in measuring the dimension of space-time
region with linear size $l$ the resolution can not be taken to a
better accuracy than $\epsilon \geq \delta l$. Since the measure
is positive definite, $\mu^{\alpha}(E,\,\epsilon) \geq
\mu^{\alpha}(E,\,0) $, the operational dimension
$\alpha_{\bf{op}}$ satisfies
\[\alpha_{\bf{op}}\, \leq \,\alpha_H~.\] In order to get
a specific value for an operational dimension one has to
generalize the definition of dimension to a finite resolution
case. Except of some "pathological" cases that have no physical
interest, the Hausdorff dimension is equivalent to the
box-counting dimension introduced by Kolmogorov \cite{Kolmogorov}.
Assume $\Omega$ is $n$ dimensional space, where $n \geq 1$, and
suppose that $N(\epsilon)$ is the minimum number of
$n$-dimensional boxes of side length $\epsilon$ required to cover
the set $E \subset \Omega$. Then the Kolmogorov (or box-counting)
dimension is defined as:
\[\dim(E) = \lim\limits_{\epsilon\rightarrow 0}{\ln
N(\epsilon)\over \ln \left(d(E) / \epsilon\right)}~,\] where
$d(E)$ is the diameter of $E$ \[d(E) = \sup\{\xi(x,\,y)\,|\, x\,,y
\in E \}~.\] The condition $E \subset \Omega$ immediately implies
$\dim(E) \leq n$. The Kolmogorov dimension can be
straightforwardly generalized to the case of a finite resolution
implying small but finite number of $\epsilon$. Thus for a
space-time region $l^4$, that is, a space-time box of side length
$l$, the operational dimension  can be defined as

\begin{equation}\label{operdim} \alpha_{\bf{op}}(l^4) =  {\ln N(\delta l)\over \ln
\left(l / \delta l\right)}~,\end{equation} where we have taken
into account finiteness of space-time resolution $\epsilon \geq
\delta l$. Now let us notice that the deviation in $\epsilon$-box
number counting can take place due to length uncertainty allowing
$l \rightarrow l - \delta l$ and correspondingly
\[ N(\delta l) = \left( {l -\delta l \over \delta l} \right)^4 ~.
\] Thus from the Eq.(\ref{operdim}) one finds the operational
dimension of $l^4$ as \[\alpha_{\bf{op}} = 4 {\ln \left( {l \over
\delta l} - 1 \right) \over \ln {l \over \delta l} } ~. \] For
large values of distance $l \gg l_P$ the uncertainty $\delta l \ll
l$ and correspondingly the deviation of space-time dimension from
four takes the form
\[\delta = 4 - \alpha_{\bf{op}} \approx {4 \delta l \over l\ln {l\over \delta l} }~.\] Denoting $r \equiv l/l_P$ for
Eqs.(\ref{Karol},\,\ref{Camelia}) one finds
\[\delta_1 \approx {6 \over r^{2/3}\ln r}~~~ \mbox{and}~~~
\delta_2 \approx {8 \over r^{1/2}\ln r}~~~\mbox{respectively}~.\]

Let us look at the experimental bounds for deviation of space-time
dimension from four considered in \cite{ZS, MS}. The experimental
bound on the deviation of space-time dimension from four coming
from the electron $\tt g-2$ factor measurement is $\sim 10^{-7}$
\cite{ZS}. At the length scale associated to this phenomenon, it
is a Compton length of the electron $r \sim 10^{22}$, the above
theoretical result gives $\delta_2 \sim 10^{-11}$. The Lamb shift
measurement for hydrogen atom puts the limit $\sim 10^{-11}$
\cite{MS}. At the atomic scale, that is, $r \sim 10^{25}$ one gets
$\delta_2 \sim 10^{-14}$. The perihelion measurement of the planet
Mercury sets the bound $\sim 10^{-9}$ \cite{MS}. But it is much
larger compared with the theoretical result even at the atomic
scale. It would be interesting if present or near future
experiments could come closer to the theoretical results. At the
nuclear scale $\sim 10^{-13}$cm, that is, $r\sim 10^{20}$ we get
$\delta_2 \sim 10^{-10}$. At the scale TeV$^{-1} \sim 10^{-17}$cm
it gives $\delta_2 \sim 10^{-8}$.

\section*{Concluding remarks}

Let us briefly summarize our discussion. The presented discussion
of space-time uncertainty relation from different points of view
strongly favors the Eq.(\ref{Camelia}). Evidently, the space-time
uncertainty relation precludes us from taking the limit $\epsilon
\rightarrow 0$ implied by the standard definitions of dimension.
Due to positive definiteness of Hausdorff measure one naturally
expects the operational dimension to be smaller than four (it is
more evident in the box-counting approach to the dimension). Even
on the basis of the above effective quantum field theory view of
the space-time uncertainty relations implying the relation between
UV and IR cutoffs one might naturally expect the space-time
dimension smaller than four. Namely, for a given IR scale the
presence of related UV cutoff implies the regularization of the
quantum field theory divergencies which equivalently well could be
done in dimensional regularization approach. Because of space-time
uncertainty relation we need to operate with a finite resolution
in estimating the space-time dimension. Box-counting dimension
introduced by Kolmogorov provides a straightforward way for an
operational definition of dimension.

Let us notice that one could use a simple physical way for
estimating of operational value of dimension. Denoting the
operational dimension of space by $3 - \delta $ one finds that the
modified Newtonian potential should behave as $V \propto r^{\delta
-1}$ ($r \equiv l/l_P $). But this effect of dimension reduction
is the result of space-time uncertainty that in the Newtonian
potential leads to the modification $r \rightarrow r - \delta r$.
So we have the same effect written in different terms. By equating
$r^{1 - \delta } = r - \delta r$ one gets \[\delta = 1 - {\ln(r -
\delta r) \over \ln r}~.\] For large values of distance $r \gg 1$
we have $\delta r \gg r$ and correspondingly this expression takes
the form $\delta \approx \delta r / r\ln r$. Thus we get the same
result in this less rigorous but physically motivated way.

The operational dimension of space-time turns out to be a scale
dependent quantity slightly smaller than four at distances $\gg
l_P$. At relatively short distances the experimental bounds
considered in \cite{ZS, MS} are $3-4$ orders of magnitude greater
than the theoretical predictions but we can hope the present or
near future experiments to approach the theoretical results.

\vspace{0.4cm}

\begin{acknowledgments}

The work was supported by the \emph{INTAS Fellowship for Young
Scientists} and the \emph{Georgian President Fellowship for Young
Scientists}.

\end{acknowledgments}

\end{document}